\documentclass{article}
\usepackage{eurosym}
\usepackage{amssymb}
\usepackage{amsfonts}
\usepackage{amsmath}
\usepackage{graphicx}
\usepackage{caption}
\usepackage{wrapfig}
\usepackage{subfig}
\usepackage{rotating}

\newtheorem{theorem}{Theorem}
\newtheorem{acknowledgement}[theorem]{Acknowledgement}

\input{tcilatex}
\begin{document}

\title{Managing COVID-19 Pandemic without Destroying the Economy\thanks{%
Original version published on arXiv on April 21, 2020.}}
\author{David Gershon,$\ $Alexander Lipton \\
%EndAName
The Jerusalem Business School, The Hebrew University of Jerusalem \  \and %
Hagai Levine \\
%EndAName
School of Public Health, The Hebrew University of Jerusalem\\
Hadassah Medical Center \ }
\maketitle

\begin{abstract}
We analyze an approach to managing the COVID-19 pandemic without shutting
down the economy while staying within the capacity of the healthcare system.
We base our analysis on a detailed heterogeneous epidemiological model,
which takes into account different population groups and phases of the
disease, including incubation, infection period, hospitalization, and
treatment in the intensive care unit (ICU). We model the healthcare capacity
as the total number of hospital and ICU beds for the whole country. We
calibrate the model parameters to data reported in several recent research
papers. For high- and low-risk population groups, we calculate the number of
total and intensive care hospitalizations, and deaths as functions of time.
The main conclusion is that countries, which enforce reasonable hygienic
measures on time can avoid lockdowns throughout the pandemic provided that
the number of spare ICU beds per million is above the threshold of about
100. In countries where the total number of ICU beds is below this
threshold, a limited period quarantine to specific high-risk groups of the
population suffices. Furthermore, in the case of an inadequate capacity of
the healthcare system, we incorporate a feedback loop and demonstrate that
quantitative impact of the lack of ICU units on the death curve. In the case
of inadequate ICU beds, full- and partial-quarantine scenarios outcomes are
almost identical, making it unnecessary to shut down the whole economy. We
also study a more detailed model with four groups, namely, children,
low-risk adults, high-risk adults, and nursing home occupants, which are
considered very high risk. We conclude that schools' opening does not
increase the risk of over-capacity of the health system or the chance for a
second wave, provided that the rest of the population behaves responsibly.
On the other hand, the risk from nursing homes to over-capacity and death
rates requires special attention to avoid a high infection rate among this
group.
\end{abstract}

\section{Introduction}

Pandemics are nothing new and have been afflicting humankind since
prehistoric times. Rather than starting with the obligatory references to
the Black Death, which devastated Europe in the fourteenth century and
killed about 30\% to 60\% of Europe's population, we briefly mention some of
the flu pandemics of the last hundred and thirty years, see \cite{CDC2020}:

(a) The \textquotedblleft Russian Flu\textquotedblright\ or
\textquotedblleft Asiatic Flu,\textquotedblright\ 1889-1890, was caused by
either H3N8 or H2N2 virus. The \textquotedblleft Russian
Flu\textquotedblright\ had a very high attack and mortality rates, with
around a million fatalities worldwide;

(b) The \textquotedblleft Spanish Flu,\textquotedblright\ 1918.1919, was
caused by the H1N1 virus and considered to be the most severe pandemic in
recent history. The \textquotedblleft Spanish Flu\textquotedblright\ had
rapidly spread on all continents to become a worldwide pandemic, which
eventually infected about 500 million people, comprising one-third of the
world's population. Being unusually deadly and virulent, it killed at least
50 million worldwide, with about 675,000 deaths occurring in the US;

(c) The \textquotedblleft Asian Flu,\textquotedblright\ 1957.58, was caused
by an H2N2 virus. Similarly to the \textquotedblleft Russian
Flu,\textquotedblright\ it caused about 1.1 million deaths worldwide and
116,000 in the US;

(d) The \textquotedblleft Hong Kong Flu,\textquotedblright\ 1968-72, was
caused by an H3N2 virus. It killed approximately one million people
worldwide and about 100,000 - in the US;

(e) The \textquotedblleft Swine Flu,\textquotedblright\ 2009.10, was caused
by an H1N1 virus and killed up to 500,000 people worldwide and approximately
13,000 in the US.

In theory, authorities can arrest an epidemic by quarantining all the
population for a prolonged period, provided that such quarantine is
technically feasible. However, the economic and social price of such
quarantine is too much to bear, not to mention its decisively medieval
nature. Expected consequences include the destruction of the economy,
enormous unemployment, and social and health aspects of quarantine, such as
isolation and loneliness, drug abuse, and domestic violence, not to mention
hunger and social unrest.

Historically, pandemics tend to attack the group, which is generally
perceived as low-risk, such as young and healthy, more than elder people,
who are usually viewed as vulnerable. For example, during the
\textquotedblleft Spanish Flu\textquotedblright\ and the \textquotedblleft
Swine Flu,\textquotedblright\ younger and healthier individuals were in
greater danger than senior citizens. In contrast, COVID-19 attacks the
elderly population much more aggressively than the younger ones, in line
with the common flu. (In all pandemics and epidemics, people with
pre-existing conditions belong to the high-risk group.) Therefore, the
response and pandemic preparedness efforts should not be primarily based on
the previously used measures but must be modeled based on the new reality on
the ground.

At the same time, the most recent pandemic of 2010 provided a lot of useful
epidemiological insights, including statistically proven the effectiveness
of face masks as an essential prevention tool. Several authors showed that
wearing face masks reduced the rate of infection for such respiratory
transmitted infections as swine flu, see, e.g., \cite{Condon2010,
Cowling2010, DelValle2010}.

Current epidemiological models are based on the simplistic assumption that
the population responds the same way to the pandemic. However, COVID-19 has
a remarkably different impact on high- and low-risk individuals. We describe
the dynamic of disease for each group, or factor, and include interactions
between them to create a close to a reality model.

In this article, we present a detailed and rich epidemiological multi-factor
model. Our model accounts for different phases of the disease, potential
aquisition of the herd immunity,multiple groups in the populations, and the
interaction between individuals belonging to various groups. Our model
treats the health system as given and accounts for such aspects as the
availability of extra hospital beds and intensive care units (ICU) to
accommodate the pent-up demand due to the pandemic. We use the most recent
research data to calibrate the model. The model allows simulating lockdowns
and quarantines of each specific group and the response of different groups
to mandatory social behavior in public.

First, we need to agree on the ultimate purpose of the lockdowns. If they
are implemented to buy time until pharmaceutical companies find a vaccine or
medical professionals introduce efficient treatments, they can potentially
protect people from dying of COVID-19. However, such an approach will lead
to economic mayhem, with many people dying from the consequences of economic
and financial destruction.

If, on the other hand, the purpose of the lockdowns is to ensure that the
pandemic spreads slowly, with a reduced epidemic peak (flattening the
curve), then a very different strategy is needed. Flattening the curve is
particularly important if the healthcare system has limited capacity
(especially in terms of medical personnel and intensive care units, ICUs.)
Such an approach can lead to reduced mortality, even if the total number of
infections remains the same.

In either case, the critical point is to set a goal and develop a strategy
of achieving it. Merely being on the safe side does not represent a viable
approach because the objective function of policymakers should include other
health issues, as well as society as a whole.

Our conclusion is predicated on the widespread implementation of sensible
pandemic response measures, such as wearing face masks, following strict
hygiene routine, social distancing, paid self-quarantine, and ongoing
surveillance via testing and possibly phone tracking. Authorities can
enforce such measures by imposing fines for violations of the regime and
help employers to provide paid leave related to quarantines.

Moreover, to reduce transmission of the \ disease, readily available
reliable testing comes to the fore. This sensible approach is especially
appropriate for such essential parts of the infrastructure, like healthcare,
police, food processing, and the likes. Fast and efficient tests improve the
detection of infected people and prevent further SARS-CoV-2 transmission,
reducing, thereby, the reproductive number, see Section \ref{SEIRK}.

Of course, testing for antibodies and issuing a clean bill of health
(provided that the virus immunity is real) to the individuals, who have
recovered from the virus, can also be of great help.

Besides, in many metropolitan areas, transportation, especially during the
rush hour, is one of the biggest, if not the biggest, source of infection.
To alleviate the transportation burden and prevent crowding during lunch
breaks, staggering work hours, extending the workweek, allowing people to
work from home on alternating days, and other similar measures, are
obviously helpful.

The main conclusion of this paper is that in countries where the majority of
the population abides by the sensible rules and regulations, the economy can
continue functioning. At the same time, the health system can expand to
manage the additional patient load. We feel that schools' opening does not
increase the risk of over-capacity of the health system or the chance for a
second wave, provided that the rest of the population behaves responsibly.
On the other hand, the risk from nursing homes to over-capacity and death
rates requires special attention to avoid a high infection rate among this
group.

The rest of the paper is organized as follows. In Section \ref{SEIRK}, we
introduce a K-group version of the basic Kermack--McKendrick model with a
generic preferred mixing matrix. We specify the model for the most important
case of two groups (the high-risk and low-risk groups) in Section \ref{SEIR2}%
. In Section \ref{4SEIR}, we consider the four-group case, the groups being
children, low-risk adults, high-risk adults, and occupants of nursing homes.
An \textit{a priori} choice of the model parameters is hard to accomplish
because the actual number of people infected by the virus is unknown due to
the prevalence of asymptomatic cases. To address this problem, we do
\textquotedblleft implied\textquotedblright\ calibration of the model by
choosing parameters in such a way that observable quantities, such as the
number of hospitalizations, including ICU admissions, and, deaths rates are
reproduced with reasonable accuracy. This calibration is described in
Section \ref{Calibration}. In Section \ref{Results} we formulate and
graphically illustrate our main results. We draw our main conclusions in
Section \ref{Conclusions}.

\section{A heterogeneous SEIR model with $K$ groups\label{SEIRK}}

The famous Kermack--McKendrick model is the prime working horse of
epidemiology, see, e.g., \cite{Anderson1992, Brauer2008}. Depending on the
particulars of the disease under consideration, one can use either
susceptible-infected-removed (SIR), or susceptible-exposed-infected-removed
(SEIR), or a more complex model. The salient features of COVID-19, however,
make it imperative to use a more advanced version of the model, which, at
the very least, should account for two groups of susceptibles - a high-risk
group, and a low-risk group, see, e.g., \cite{Choe2015}. This separation is
a must because the average mortality rate, which is of the order of $0.4$
percent or so, see, \cite{Lipton2020}, is not a good representation of
actual mortality, which is manifestly bimodal. Besides, the model should
involve the all-important feedback loop, which accounts for the fact that
mortality increases when the demand for intensive care outstrips its supply
so that rationing becomes inevitable and results in elevated mortality.

We separate the population in $K$ groups. Let $N_{k}$ be the number of
people in the $k$-th group, and $a_{k}$ is the average number of contacts
per person per unit of time. Conditional on data availability, this approach
allows us to model the pandemic on a more granular level, accounting for
varying vulnerability levels, different population densities in metropolitan
areas within a country or different areas within a given city, demographics,
and the likes.

The critical assumption is that the number of contacts between members of
the $k$-th and $k^{\prime }$-th group, per unit of time is given by%
\begin{equation}
\begin{array}{c}
C_{kk^{\prime }}=\left( \pi _{k}\delta _{kk^{\prime }}+\left( 1-\pi
_{k}\right) \frac{\left( 1-\pi _{k^{\prime }}\right) a_{k^{\prime
}}N_{k^{\prime }}}{\dsum\nolimits_{k^{\prime }=1}^{K}\left( 1-\pi
_{k^{\prime }}\right) a_{k^{\prime }}N_{k^{\prime }}}\right) a_{k}N_{k},%
\end{array}
\label{Eq1}
\end{equation}%
where $\delta _{kk^{\prime }}$ is the Kronecker symbol. Here $\pi =\left(
\pi _{1},...,\pi _{K}\right) $, $0\leq \pi _{k}\leq 1$, is a vector, which
characterizes the relative size of interactions within and outside of the $k$%
-th group. The matrix $C=\left( C_{kl}\right) $ has rank one when $\pi
_{k}=0 $, which corresponds to proportional mixing, while it has a diagonal
form for $\pi _{k}=1$, which corresponds to no mixing between the groups,
caused, for example, by cultural preferences, or a lockdown. The matrix $C$
is called the preferred mixing matrix.

A given group is subdivided into susceptibles, exposed, infected,
hospitalized, moved to the intesive care unit (and put on ventilator),
recovered, and dead, $S_{k}$, $E_{k}$, $I_{k}$, $H_{k}$, $V_{k}$, $R_{k}$, $%
D_{k}$, respectively. Such classification corresponds to different phases of
the disease. In what follows, we use the data points from \cite%
{Ferguson2020, Kissler2020}.

The population in the $k$-th class, which is not infected at time $t$ forms
the susceptible group; with the number of susceptible individuals denoted by 
$S_{k}$. The first phase of the disease starts when a susceptible individual
becomes exposed via a contact with an infectious individual, but does not
show symptoms and cannot yet infect others. We assume that, on average, a
representative individual is exposed for $\tau _{E}$ days until she can
infect others. Typically, $\tau _{E}=4.6$ days. We denote the number of
exposed people in the $k$-th class at time $t$ by $E_{k}(t)$.

The next phase begins when an exposed individual becomes infected. We denote
the number of infected individuals in the $k$-th class k as $I_{k}(t)$.
Every exposed individual stays infected $\tau _{I}$ days. Typically, $\tau
_{I}=4$ days. At this stage, infected people transmit the disease,
regardless of exhibiting symptoms or being asymptomatic.

During the next phase, a certain percentage of the infected individuals
requires hospitalizations, depending on their risk group. Without
hospitalization, the probability of recovery or death is $\beta _{k}^{\left(
R\right) }$ and $\beta _{k}^{\left( D\right) }$, respectively. We denote the
number of hospitalized individuals in the $k$-th class at time $t$ as $%
H_{k}(t)$ and the probability of hospitalization $\beta _{k}^{\left(
H\right) }$. On average, an individual stays in the hospital for $\tau _{H}$
days before they are either moved to the ICU or recover. Typically, $\tau
_{H}=7$ days. The probability of recovery or death without being moved to
intensive care is $\gamma _{k}^{\left( R\right) }$ and $\gamma _{k}^{\left(
D\right) }$, respectively.

A certain percentage of severely ill hospitalized individuals must be moved
to an ICU; this percentage depends on the class (risk group) they belong to.
The number of individuals from the $k$-th class staying at the ICU is $%
V_{k}(t)$ and the probability that a hospitalized individual requires the
ICU treatment is $\gamma _{k}^{\left( V\right) }$.

The average time in the ICU is $\tau _{V}$ days. Typically, $\tau _{V}=10$
days. For individuals receiving the ICU treatment, the probability to return
to the regular hospital is $\delta _{k}^{\left( H\right) }$, and the
probability of dying is $\delta _{k}^{\left( D\right) }$, respectively. We
denote by $R_{k}(t)$ and $D_{k}(t)$ the numbers of individuals in the $k$-th
class who have recovered or died until time $t$, respectively.

The corresponding ODEs have the form%
\begin{equation}
\begin{array}{c}
\frac{dS_{k}}{dt}=-a_{k}\Theta _{k}S_{k}, \\ 
\\ 
\frac{dE_{k}}{dt}=a_{k}\Theta _{k}S_{k}-\frac{1}{\tau _{E}}E_{k}, \\ 
\\ 
\frac{dI_{k}}{dt}=\frac{1}{\tau _{E}}E_{k}-\frac{1}{\tau _{I}}I_{k}, \\ 
\\ 
\frac{dH_{k}}{dt}=\frac{\beta _{k}^{\left( H\right) }}{\tau _{I}}I_{k}+\frac{%
\delta _{k}^{\left( H\right) }}{\tau _{V}}V_{k}-\frac{1}{\tau _{H}}H_{k}, \\ 
\\ 
\frac{dV_{k}}{dt}=\frac{\gamma ^{\left( V\right) }}{\tau _{V}}H_{k}-\frac{1}{%
\tau _{V}}V_{k}, \\ 
\\ 
\frac{dR_{k}}{dt}=\frac{\beta _{k}^{\left( R\right) }}{\tau _{I}}I_{k}+\frac{%
\gamma _{k}^{\left( R\right) }}{\tau _{H}}H_{k}, \\ 
\\ 
\frac{dD_{k}}{dt}=\frac{\beta _{k}^{\left( D\right) }}{\tau _{I}}I_{k}+\frac{%
\gamma _{k}^{\left( D\right) }}{\tau _{H}}H_{k}+\frac{\delta _{k}^{\left(
D\right) }}{\tau _{V}}V_{k}, \\ 
\\ 
\frac{dN_{k}}{dt}=-\left( \frac{\beta _{k}^{\left( D\right) }}{\tau _{I}}%
I_{k}+\frac{\gamma _{k}^{\left( D\right) }}{\tau _{H}}H_{k}+\frac{\delta
_{k}^{\left( D\right) }}{\tau _{V}}V_{k}\right) .%
\end{array}
\label{Eq3}
\end{equation}%
Here the transitional coefficients are inversely proportional to the average
time spent by a representative individual in different states, $\tau _{E}$, $%
\tau _{I}$, $\tau _{H}$, $\tau _{V}$.

The all-important quantity in Eqs (\ref{Eq3})$\Theta _{k}$ depends on vector 
$\pi =\left( \pi _{1},...,\pi _{K}\right) $, which characterizes the
intensity of interactions for the members of the $k$-th group and is given by%
\begin{equation}
\begin{array}{c}
\Theta _{k}=\pi _{k}\frac{I_{k}}{N_{k}}+\left( 1-\pi _{k}\right) \Omega , \\ 
\\ 
\Omega =\frac{\dsum\nolimits_{k^{\prime }=1}^{K}\left( 1-\pi _{k^{\prime
}}\right) a_{k^{\prime }}I_{k^{\prime }}}{\dsum\nolimits_{k^{\prime
}=1}^{K}\left( 1-\pi _{k^{\prime }}\right) a_{k^{\prime }}N_{k^{\prime }}}.%
\end{array}
\label{Eq4}
\end{equation}%
It is chosen to be in agreement with Eq. (\ref{Eq1}).

By construction, Eqs (\ref{Eq3}) satisfy the conservation laws%
\begin{equation}
\begin{array}{c}
\frac{d}{dt}\left( S_{k}+E_{k}+I_{k}+H_{k}+V_{k}+R_{k}-N_{k}\right) =0, \\ 
\\ 
\frac{d}{dt}\left( D_{k}+N_{k}\right) =0.%
\end{array}
\label{Eq8}
\end{equation}

Eqs (\ref{Eq3}) have to be supplied with the initial conditions.\ We choose
these conditions in the form%
\begin{equation}
\begin{array}{c}
S_{k}\left( 0\right) =\left( 1-\xi _{k}\right) N_{k},\ \ \ E_{k}\left(
0\right) =\xi _{k}N_{k},\ \ \ I_{k}\left( 0\right) =0,\ \ \ H_{k}\left(
0\right) =0, \\ 
V_{k}\left( 0\right) =0,\ \ \ R_{k}\left( 0\right) =0,\ \ \ D_{k}\left(
0\right) =0,\ \ \ N_{k}\left( 0\right) =N_{k},%
\end{array}
\label{Eq8a}
\end{equation}%
where $N_{k}$ is the initial number of individuals in their respecitve
class, while $\xi _{k}$ is the fraction of the population \emph{initially}
exposed to the virus. In principle, it can be as small as $1/N_{k}$ - the
proverbial \textquotedblleft patient zero.\textquotedblright

\begin{figure}[th]
{\center\includegraphics[width=1.0\textwidth, angle=0]
{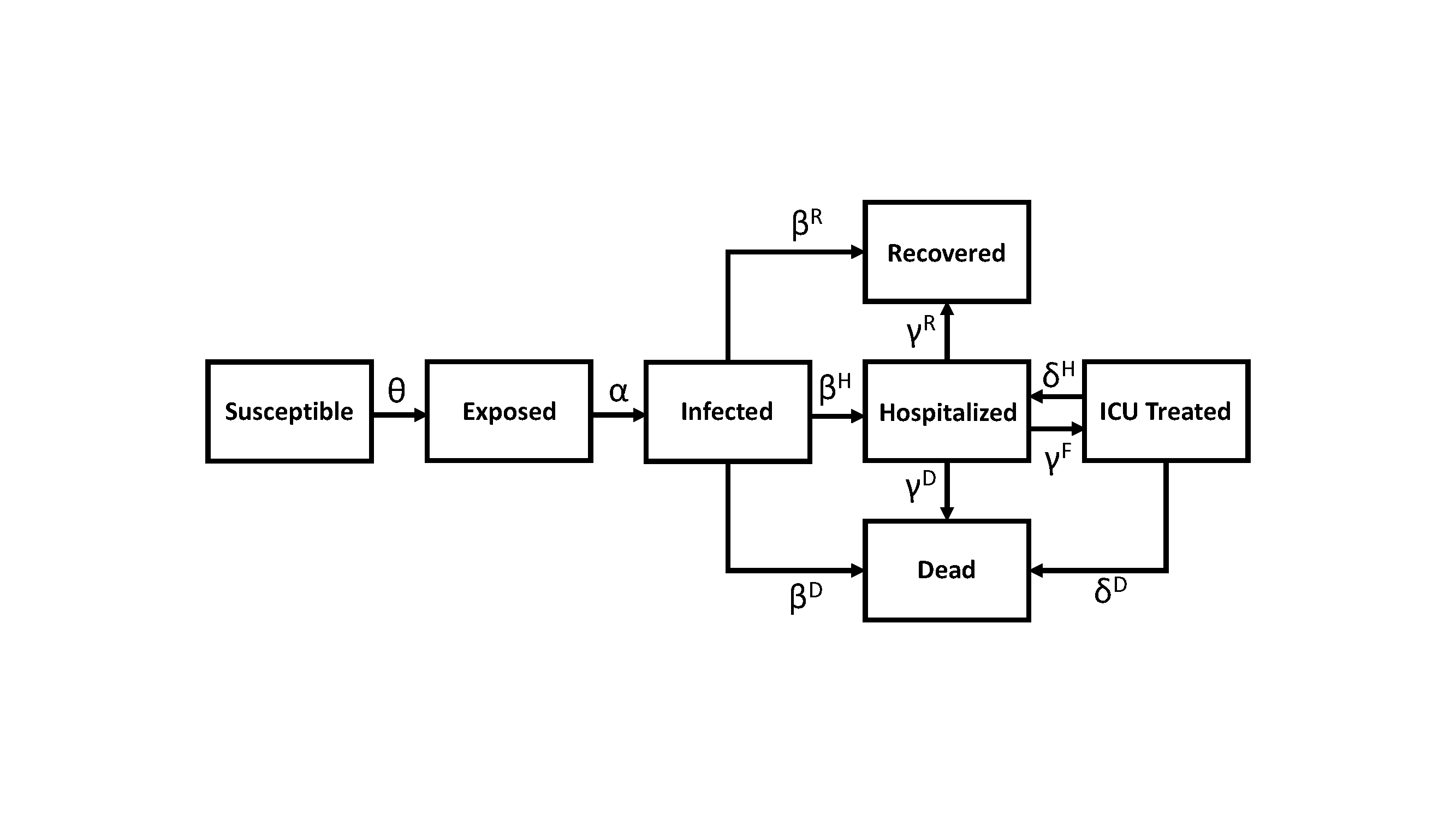} }
\caption{{}Flow chart for the SEIR model.}
\label{Fig1}
\end{figure}

The flowchart for Eqs (\ref{Eq3}) is shown in Figure \ref{Fig1}.%
\begin{equation*}
\text{Figure \ref{Fig1}\ near here.}
\end{equation*}

We note in passing that by choosing values independent of $k$ for all the
relevant quantities and putting $\pi _{k}=0$, we can convert the model into
the particular case of the single group.

According to \cite{Ferguson2020, Kissler2020}, the rough estimates for
timing parameters are as follows%
\begin{equation}
\begin{array}{c}
\tau _{E}=4.6\text{ days},\ \ \ \tau _{I}=4\text{ days},\ \ \ \tau _{H}=7%
\text{ days},\ \ \ \tau _{V}=10\text{ days}.%
\end{array}
\label{Eq3c}
\end{equation}%
The actual process in assigning population to varous risk groups is
relatively involved. We discuss it in the Section \ref{SEIR2} for the most
relevant and practical case of high- and low-risk groups, $K=2$.

We note in passing that the way Eqs (\ref{Eq3}), (\ref{Eq4}) are manifestly
scale-invariant. In other words, the increase in the overall population
leads to a proportional increase in the number of beds and deaths.However,
in large metropolitan centers, it is more appropriate to choose $\Theta _{k}$
in the form which violates scaling symmetry:%
\begin{equation}
\begin{array}{c}
\Theta _{k}^{\left( \omega \right) }=\pi _{k}\frac{I_{k}}{N_{k}^{1-\varpi }}%
+\left( 1-\pi _{k}\right) \Omega ^{\left( \varpi \right) }, \\ 
\\ 
\Omega ^{\left( \varpi \right) }=\frac{\dsum\nolimits_{k^{\prime
}=1}^{K}\left( 1-\pi _{k^{\prime }}\right) a_{k^{\prime }}I_{k^{\prime }}}{%
\dsum\nolimits_{k^{\prime }=1}^{K}\left( 1-\pi _{k^{\prime }}\right)
a_{k^{\prime }}N_{k^{\prime }}^{1-\varpi }},%
\end{array}
\label{Eq9}
\end{equation}%
where $\varpi $ is the factor accounting for the size of the metropolitan
area of interest and modes of human interactions that are specific to large
urban areas, such as an extensive public transportation system. A detailed
analysis will be presented elsewhere.

It is critical to realize that the ICU capacity is finite, and model it as
such, hence 
\begin{equation}
\begin{array}{c}
\delta _{k}^{\left( H\right) }=\bar{\delta}_{k}^{\left( H\right) }\mathcal{H}%
\left( C-\dsum\nolimits_{k^{\prime }=1}^{K}V_{k^{\prime }}\right) , \\ 
\\ 
\delta _{k}^{\left( D\right) }=\bar{\delta}_{k}^{\left( D\right) }+\bar{%
\delta}_{k}^{\left( H\right) }\left( 1-\mathcal{H}\left(
C-\dsum\nolimits_{k^{\prime }=1}^{K}V_{k^{\prime }}\right) \right) .%
\end{array}
\label{Eq5}
\end{equation}%
where $\mathcal{H}\left( .\right) $ is the Heaviside step function. The
celebrated reproductive number $\rho _{k}$ is defined as%
\begin{equation}
\begin{array}{c}
\rho _{k}=\tau _{I}a_{k},%
\end{array}
\label{Eq11}
\end{equation}%
This number represents a measure of transmissibility but has no clear
theoretical meaning. Therefore, instead, we use common sense and calibrate
it to the available data. Eq. (\ref{Eq11}) shows that reproductive number
for the $k$-th class is proportional to the number of contacts per unit of
time, $a_{k}$ over the duration of the infectious stage. In other words, $%
a_{k}$ represents the tendency of infected individuals from the $k$-th class
to infect other individuals either from their own class or other classes.

We model the lockdown by the proportional reduction of the coefficients $%
\rho _{k}\left( t\right) $, where the time-dependence represents seasonality
and potential switching of lockdowns on and off:%
\begin{equation}
\begin{array}{c}
\rho _{k}\left( t\right) =\phi _{k}\left( t\right) \chi _{k}\left( t\right)
\rho _{k}^{\left( 0\right) },%
\end{array}
\label{Eq6}
\end{equation}%
with%
\begin{equation}
\begin{array}{c}
\phi _{k}\left( t\right) =\dsum_{l=1}^{L}\bar{\phi}_{k,l}\mathcal{H}\left(
\left( t-t_{l-1}\right) \left( t_{l}-t\right) \right) , \\ 
\\ 
\chi _{k}\left( t\right) =1+\vartheta _{k}\cos \left( \frac{2\pi t}{T}%
\right) .%
\end{array}
\label{Eq7}
\end{equation}%
Here we assume that $\rho _{k}^{\left( 0\right) }$ is the \textquotedblleft
natural\textquotedblright\ reproductive number specific to COVID-19, $%
\vartheta _{k}$ corresponds to its seasonal variations, $T_{0}=0$, $%
T_{1}...T_{L-1}$, $T_{L}=T$, $T$ is the terminal time for the calculation,
e.g. $T=365$ days, and $T_{1}...T_{L-1}$ are times when lockdown policies
are changed. In the simplest case of a one-time lockdown, we have $L=3$; $%
T_{1}$ is the time when the quarantine is imposed, $T_{2}$ is the time when
it is lifted, $T_{3}=T$. More generally, there can be several rounds of
switching the quarantine policies on and off.

Expressed in terms of $\rho _{k}\left( t\right) $, Eqs (\ref{Eq4}), (\ref%
{Eq11}) become%
\begin{equation}
\begin{array}{c}
\Theta _{k}=\pi _{k}\frac{I_{k}}{N_{k}}+\left( 1-\pi _{k}\right) \Omega , \\ 
\\ 
\Omega =\frac{\dsum\nolimits_{k^{\prime }=1}^{K}\left( 1-\pi _{k^{\prime
}}\right) \rho _{k^{\prime }}\left( t\right) I_{k^{\prime }}}{%
\dsum\nolimits_{k^{\prime }=1}^{K}\left( 1-\pi _{k^{\prime }}\right) \rho
_{k^{\prime }}\left( t\right) N_{k^{\prime }}},%
\end{array}
\label{Eq6a}
\end{equation}%
\begin{equation}
\begin{array}{c}
\Theta _{k}^{\left( \varpi \right) }=\pi _{k}\frac{I_{k}}{N_{k}^{1-\varpi }}%
+\left( 1-\pi _{k}\right) \Omega ^{\left( \varpi \right) }, \\ 
\\ 
\Omega ^{\left( \varpi \right) }=\frac{\dsum\nolimits_{k^{\prime
}=1}^{K}\left( 1-\pi _{k^{\prime }}\right) \rho _{k^{\prime }}\left(
t\right) I_{k^{\prime }}}{\dsum\nolimits_{k^{\prime }=1}^{K}\left( 1-\pi
_{k^{\prime }}\right) \rho _{k^{\prime }}\left( t\right) N_{k^{\prime
}}^{1-\varpi }}.%
\end{array}
\label{Eq6b}
\end{equation}

Assuming that the reproductive number is constant, $\rho _{k}\left( t\right)
=\bar{\rho}_{k}$, so that there is no seasonality and no interventions, we
can calculate the asymptotic fractions of the population, which is
unaffected by the pandemic by the time it runs its course. To put it
differently, this is the level at which the herd immunity plays its magic.
Let $s_{k}=S_{k}\left( \infty \right) /S_{k}\left( 0\right) $. When $k=1$
(one group), it is a well-known fact that $s_{1}$ satisfies the following
transcendent equation%
\begin{equation}
\begin{array}{c}
\ln \left( s_{1}\right) +\bar{\rho}_{1}\left( 1-s_{1}\right) =0,%
\end{array}
\label{Eq7a}
\end{equation}%
see, e.g., \cite{Anderson1992, Brauer2008}. When $k=2$ (two groups), a
similar result can be found in \cite{Choe2015}, see below. Finally, in the
general case, the transcendent equations for the vector $s=\left(
s_{1},...,s_{K}\right) $ have the form:%
\begin{equation}
\begin{array}{c}
\ln \left( s_{k}\right) +\bar{\rho}_{k}\dsum\nolimits_{k^{\prime
}=1}^{K}\eta _{kk^{\prime }}\left( 1-s_{k^{\prime }}\right) =0,%
\end{array}
\label{Eq7b}
\end{equation}%
where%
\begin{equation}
\begin{array}{c}
\eta _{kk^{\prime }}=\pi _{k}\delta _{kk^{\prime }}+\left( 1-\pi _{k}\right) 
\frac{\rho _{k^{\prime }}\tilde{N}_{k^{\prime }}}{\dsum\nolimits_{k^{\prime
\prime }=1}^{K}\rho _{k^{\prime \prime }}\tilde{N}_{k^{\prime \prime }}}, \\ 
\tilde{N}_{k}=\left( 1-\pi _{k}\right) N_{k}.%
\end{array}
\label{Eq7c}
\end{equation}%
The asymptotic values $\left( s_{1},...,s_{K}\right) $ can be used to get an
estimate of the total number of deaths for a given set of reproductive
numbers $\left( \bar{\rho}_{1},...,\bar{\rho}_{K}\right) $ and mortality
rates $\left( \mu _{1},...,\mu _{K}\right) $, provided that authorities do
not apply mitigating measures:%
\begin{equation}
\begin{array}{c}
D_{k}=\mu _{k}\left( 1-s_{k}\right) N_{k}.%
\end{array}
\label{Eq7d}
\end{equation}

\section{The 2-SEIR model\label{SEIR2}}

For our purposes, it is natural to divide the population into high- and
low-risk groups defined by their reaction to infection, $K=2$. These groups
have completely different probabilities of being exposed, hospitalized,
moved to the ICU, and dying. The two groups are healthy (low-risk, $LR$, $%
k=1 $) and vulnerable (high-risk, $HR$, $k=2$) populations. Age is an
important, but not the only, determinant of the group. Other factors, such
as pre-exising conditions, obesity, etc. are also essential.

In the two-group case, original Eqs (\ref{Eq3}) can be simplified as follows%
\begin{equation}
\begin{array}{c}
\frac{dS_{1}}{dt}=-\frac{\rho _{1}}{\tau _{I}}\Theta _{1}S_{1}, \\ 
\\ 
\frac{dE_{1}}{dt}=\frac{\rho _{1}}{\tau _{I}}\Theta _{1}S_{1}-\frac{1}{\tau
_{E}}E_{1}, \\ 
\\ 
\frac{dI_{1}}{dt}=\frac{1}{\tau _{E}}E_{1}-\frac{1}{\tau _{I}}I_{1}, \\ 
\\ 
\frac{dH_{1}}{dt}=\frac{\beta _{1}^{\left( H\right) }}{\tau _{I}}I_{1}+\frac{%
\delta _{1}^{\left( H\right) }}{\tau _{V}}V_{1}-\frac{1}{\tau _{H}}H_{1}, \\ 
\\ 
\frac{dV_{1}}{dt}=\frac{\gamma _{1}^{\left( V\right) }}{\tau _{H}}H_{1}-%
\frac{1}{\tau _{V}}V_{1}, \\ 
\\ 
\frac{dR_{1}}{dt}=\frac{\beta _{1}^{\left( R\right) }}{\tau _{I}}I_{1}+\frac{%
\gamma _{1}^{\left( R\right) }}{\tau _{H}}H_{1}, \\ 
\\ 
\frac{dD_{1}}{dt}=\frac{\beta _{1}^{\left( D\right) }}{\tau _{I}}I_{1}+\frac{%
\gamma _{1}^{\left( D\right) }}{\tau _{H}}H_{1}+\frac{\delta _{1}^{\left(
D\right) }}{\tau _{V}}V_{1}, \\ 
\\ 
\frac{dN_{1}}{dt}=-\left( \frac{\beta _{1}^{\left( D\right) }}{\tau _{I}}%
I_{1}+\frac{\gamma _{1}^{\left( D\right) }}{\tau _{H}}H_{1}+\frac{\delta
_{1}^{\left( D\right) }}{\tau _{V}}V_{1}\right) ,%
\end{array}
\label{Eq13}
\end{equation}%
\begin{equation}
\begin{array}{c}
\frac{dS_{2}}{dt}=-\frac{\rho _{2}}{\tau _{I}}\Theta _{2}S_{2}, \\ 
\\ 
\frac{dE_{2}}{dt}=\frac{\rho _{2}}{\tau _{I}}\Theta _{2}S_{2}-\frac{1}{\tau
_{E}}E_{2}, \\ 
\\ 
\frac{dI_{2}}{dt}=\frac{1}{\tau _{E}}E_{2}-\frac{1}{\tau _{I}}I_{2}, \\ 
\\ 
\frac{dH_{2}}{dt}=\frac{\beta _{2}^{\left( H\right) }}{\tau _{I}}I_{2}+\frac{%
\delta _{2}^{\left( H\right) }}{\tau _{V}}V_{2}-\frac{1}{\tau _{H}}H_{2}, \\ 
\\ 
\frac{dV_{2}}{dt}=\frac{\gamma _{2}^{\left( V\right) }}{\tau _{H}}H_{2}-%
\frac{1}{\tau _{V}}V_{2}, \\ 
\\ 
\frac{dR_{2}}{dt}=\frac{\beta _{2}^{\left( R\right) }}{\tau _{I}}I_{2}+\frac{%
\gamma _{2}^{\left( R\right) }}{\tau _{H}}H_{2}, \\ 
\\ 
\frac{dD_{2}}{dt}=\frac{\beta _{2}^{\left( D\right) }}{\tau _{I}}I_{2}+\frac{%
\gamma _{2}^{\left( D\right) }}{\tau _{H}}H_{2}+\frac{\delta _{2}^{\left(
D\right) }}{\tau _{V}}V_{2}, \\ 
\\ 
\frac{dN_{2}}{dt}=-\left( \frac{\beta _{2}^{\left( D\right) }}{\tau _{I}}%
I_{2}+\frac{\gamma _{2}^{\left( D\right) }}{\tau _{H}}H_{2}+\frac{\delta
_{2}^{\left( D\right) }}{\tau _{V}}V_{2}\right) ,%
\end{array}
\label{Eq14}
\end{equation}%
where%
\begin{equation}
\begin{array}{c}
\Theta _{1}=\pi _{1}\frac{I_{1}}{N_{1}}+\left( 1-\pi _{1}\right) \Omega , \\ 
\\ 
\Theta _{2}=\pi _{2}\frac{I_{2}}{N_{2}}+\left( 1-\pi _{2}\right) \Omega , \\ 
\\ 
\Omega =\frac{\left( 1-\pi _{1}\right) a_{1}I_{1}+\left( 1-\pi _{2}\right)
a_{2}I_{2}}{\left( 1-\pi _{1}\right) a_{1}N_{1}+\left( 1-\pi _{2}\right)
a_{2}N_{2}}%
\end{array}
\label{Eq15}
\end{equation}%
\begin{equation}
\begin{array}{c}
\delta _{1}^{\left( H\right) }=\bar{\delta}_{1}^{\left( H\right) }\mathcal{H}%
\left( C-V_{1}-V_{2}\right) ,\ \ \ \delta _{1}^{\left( D\right) }=\bar{\delta%
}_{1}^{\left( D\right) }+\bar{\delta}_{1}^{\left( H\right) }\left( 1-%
\mathcal{H}\left( C-V_{1}-V_{2}\right) \right) , \\ 
\\ 
\delta _{2}^{\left( R\right) }=\bar{\delta}_{2}^{\left( H\right) }\mathcal{H}%
\left( C-V_{1}-V_{2}\right) ,\ \ \ \delta _{2}^{\left( D\right) }=\bar{\delta%
}_{2}^{\left( D\right) }+\bar{\delta}_{2}^{\left( H\right) }\left( 1-%
\mathcal{H}\left( C-V_{1}-V_{2}\right) \right) .%
\end{array}
\label{Eq16}
\end{equation}%
Thus, Eqs (\ref{Eq13}), (\ref{Eq14}) are connected via $\Theta _{1}$, $%
\Theta _{2}$, and $V=V_{1}+V_{2}$.

\begin{sidewaysfigure}[th]
{\center}%
\includegraphics[width=1.0\textwidth, height=1.0\textheight, angle=0]
{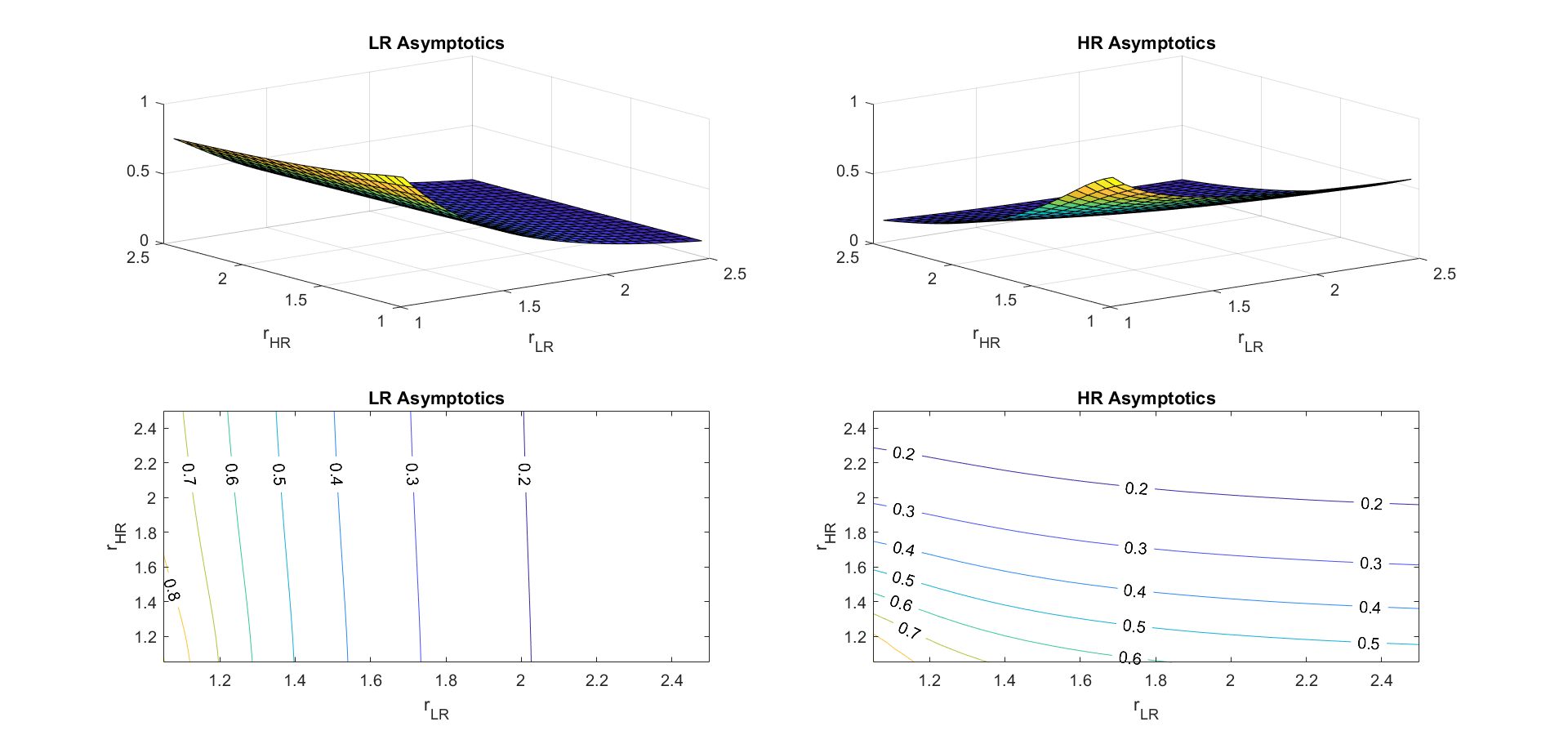}
\caption{{}Fractions of the LR and HR susceptibles not infected in the course of pandemics as functions of the reproductive rates $R_{L}$ and $R_{H}$..}
\label{Fig2}
\end{sidewaysfigure}

The two-factor version of Eq. (\ref{Eq7b}), which can be used to calculate
the asymptotic fractions of the population, which is unaffected by the
pandemic has the form:%
\begin{equation}
\begin{array}{c}
\left\{ 
\begin{array}{c}
\ln \left( s_{1}\right) +\bar{\rho}_{1}\left( \left( \pi _{1}+\frac{\left(
1-\pi _{1}\right) \bar{\rho}_{1}\tilde{N}_{1}}{\bar{\rho}_{1}\tilde{N}_{1}+%
\bar{\rho}_{2}\tilde{N}_{2}}\right) \left( 1-s_{1}\right) +\frac{\left(
1-\pi _{1}\right) \bar{\rho}_{2}\tilde{N}_{2}}{\bar{\rho}_{1}\tilde{N}_{1}+%
\bar{\rho}_{2}\tilde{N}_{2}}\left( 1-s_{2}\right) \right) =0, \\ 
\ln \left( s_{2}\right) +\bar{\rho}_{2}\left( \frac{\left( 1-\pi _{2}\right) 
\bar{\rho}_{1}\tilde{N}_{1}}{\bar{\rho}_{1}\tilde{N}_{1}+\bar{\rho}_{2}%
\tilde{N}_{2}}\left( 1-s_{1}\right) +\left( \pi _{2}+\frac{\left( 1-\pi
_{2}\right) \bar{\rho}_{2}\tilde{N}_{2}}{\bar{\rho}_{1}\tilde{N}_{1}+\bar{%
\rho}_{2}\tilde{N}_{2}}\right) \left( 1-s_{2}\right) \right) =0,%
\end{array}%
\right.%
\end{array}
\label{Eq16a}
\end{equation}%
where $\tilde{N}_{k}=\left( 1-\pi _{k}\right) N_{k}$, $k=1,2$. We show $%
\left( s_{1},s_{2}\right) $ as functions of $\left( \bar{\rho}_{1},\bar{\rho}%
_{2}\right) $ for representative $\left( \pi _{1}=0.9,\pi _{2}=0.7\right) $
in Figure \ref{Fig2}.%
\begin{equation*}
\text{Figure \ref{Fig2}\ near here.}
\end{equation*}

\section{The 4-SIER Model\label{4SEIR}}

The 2-group model we presented in the previous section improves the
predictability of the pandemic dramatically compared to the simplistic
one-group model. However, in many countries, it is imperative to include
further groups and increase the granularity of the population in the model
to improve the efficacy of the model. One has to choose the number of groups
depending on the characteristics of the disease under consideration. In the
case of COVID-19 naturally, there are two other groups to consider: Children
and population in nursery/retirement homes. Children are at a lower risk of
severe morbidity mortality. Several researchers argued that children are
typically less infectious than adults. Based on many countries' experience,
the disease spreads out extremely fast in nursery/retirement homes, which
are semi-confined groups and yet include in its majority high or even very
high-risk population. One can extract a lot of information about the NH
group, since a high proportion of the people consuming ICU beds and dying
come from the nursery/retirement homes. Hence with a 4-group model, we can
also improve the partial quarantine strategy to include more specific
populations and calculate the effect of closing and opening schools.

Our 4-group model are as follows:

\begin{enumerate}
\item Children with no chronic illnesses ($C$).

\item Adults aged less than 65 years without chronic diseases, low risk ($LR$%
).

\item Adults aged 65 years and all the younger population with chronic
illnesses, high risk ($HR$).

\item People in nursing/retirement homes ($NH$).
\end{enumerate}

Typically, the ratio of this population is known roughly in every country.
We denote the four groups by $C,LR,HR,NH$.

Going to eq. (2), now $k=1,2,3,4$. Ostensibly, we need to calibrate twice as
many parameters as before and find $\beta (k),\gamma (k),\delta (k)$ for $%
k=1,2,3,4$, instead of $k=1,2$. Fortunately, this is not exactly the case.
We know that children's hospitalizations are exceedingly rare, while their
death rate is almost null. We can get some indication of the parameters for
nursing/retirement homes from the \textquotedblleft
Corona\textquotedblright\ boat that docked in Japan for a few weeks. The
calibration allows us actually to estimate the infection rate of children
since, in most countries, they are rarely tested and seldomly hospitalized.
If we, therefore, assume that the national measured reproduction rate most
likely ignores children, then the reproduction rate of children as a group
(e.g., in schools and nursery gardens) can be approximated quite accurately.

To demonstrate the power of the 4-group model in decision making, let us
consider a small country like Israel, with a population of 9.1 million
people. The four groups, $k=1,2,3,4$, are:

\begin{enumerate}
\item $k=1$ - 1 million Children with no background diseases.

\item $k=2$ - 6.5 million Adults with no background diseases with age less
than 65.

\item $k=3$ - 1.5 million Elderly population above age 65 and young
population with background diseases.

\item $k=4$ - 100,000 People in nursery/retirement homes.
\end{enumerate}

While the $NH$ group is only $10\%$ of the size of the $HR$ group, it
affects the ICU consumption significantly, because, being a closed place,
nursing home inhabitants have a much higher reproductive rate than the
general population, unless nursing homes are thoroughly isolated.

The $C$ group has a significant effect as well since it is a `hidden' group
in terms of hospitalization but might affect the spread-out of the disease
significantly.

\section{Calibration\label{Calibration}}

We calibrate the model to recent clinical data and mortality statistics, 
\cite{Ferguson2020, Kissler2020, World2020}. The idea is that some
parameters of the model, such as the reproductive rates $\rho $, and the
specific times $\tau _{E}\,$(possibly), $\tau _{I}$, $\tau _{H}$, $\tau _{V}$%
, can be observed from clinical observations. This is also true for $\gamma $
and $\delta $. Finding $\beta $ is more difficult because its proper
evaluation is predicated on knowing the number of infected individuals,
which is unobservable. Our strategy is to calibrate it to observables, such
as the number of hospitalizations, ICU hospitalizations, and deaths. For
example, given our previous discussion, we can choose the coefficients$\
\beta ,\gamma ,\delta ,\pi $ for Sweden as shown in Table \ref{Tab1}.%
\begin{equation*}
\text{Table \ref{Tab1}\ near hear.}
\end{equation*}%
Here $1\rightarrow LR$, $2\rightarrow HR$. This choice results in the
following mortality rates for the $LR$ and $HR$ groups $p_{1,2}^{\left(
D\right) }$ and $p^{\left( D\right) }$ for the population as a whole:%
\begin{equation}
\begin{array}{c}
p_{1}^{\left( D\right) }=0.83\%,\ \ \ p_{2}^{\left( D\right) }=1.01\%,\ \ \
p^{\left( D\right) }=0.21\%.%
\end{array}
\label{Eq19}
\end{equation}%
Although these estimates are a little lower than some of the current
projections, they fit the actual pattern on the ground well.

For the four-group case, we choose parameters as shown in Table \ref{Tab2}.%
\begin{equation*}
\text{Table \ref{Tab2} near here.}
\end{equation*}%
Here $1\rightarrow C$, $2\rightarrow LR$, $3\rightarrow HR$, $4\rightarrow
NH $.

\section{Results\label{Results}}

In this section, we use parameters given in Table \ref{Tab1} with the
reproductive numbers $\rho _{1}\left( t\right) $, $\rho _{2}\left( t\right) $
corresponding to the situation in Sweden and in Israel, and analyze several
different strategies for dealing with the pandemic.

\begin{sidewaysfigure}[th]
{\center}%
\includegraphics[width=1.0\textwidth, height=1.0\textheight, angle=0]
{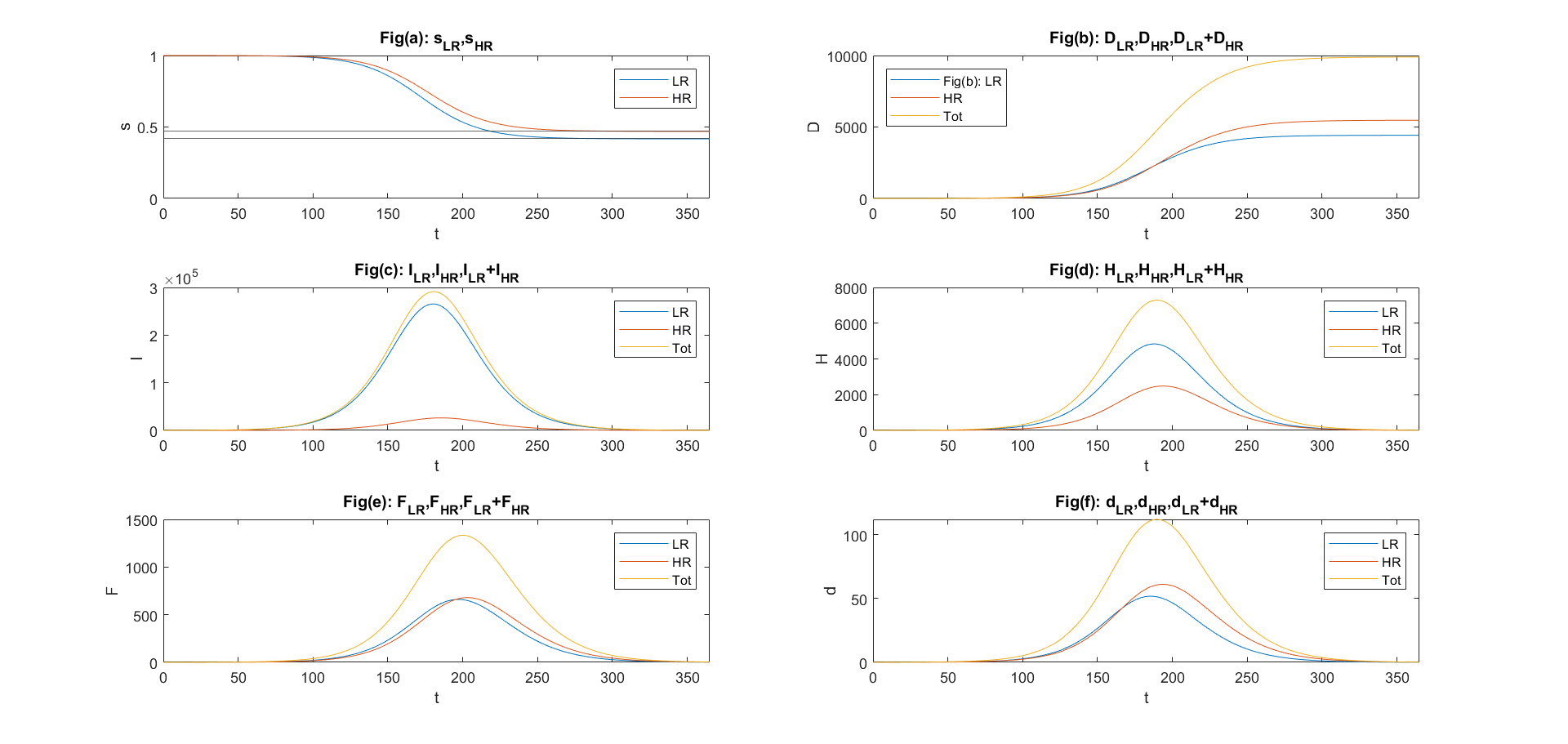}
\caption{{}Disease dynamics, in Sweden, sufficient number of ICU beds, without seasonality.}
\label{Fig3A}
\end{sidewaysfigure}

In Figure \ref{Fig3A}, we show the dynamic of the disease in Sweden
corresponding to parameters givenin Table \ref{Tab1} and%
\begin{equation}
\begin{array}{c}
\rho _{1}=1.50,\ \ \ \rho _{2}=1.40,%
\end{array}
\label{Eq19a}
\end{equation}%
so that there is no seasonality. It is clear from Figure \ref{Fig3A}(a) that
the asymptotic fractions of the population, which is unaffected by the
pandemic, given by Eqs (\ref{Eq16a}) agree with our calculations entirely.

\begin{equation*}
\text{Figure \ref{Fig3A} near here.}
\end{equation*}

\begin{sidewaysfigure}[th]
{\center}%
\includegraphics[width=1.0\textwidth, height=1.0\textheight, angle=0]
{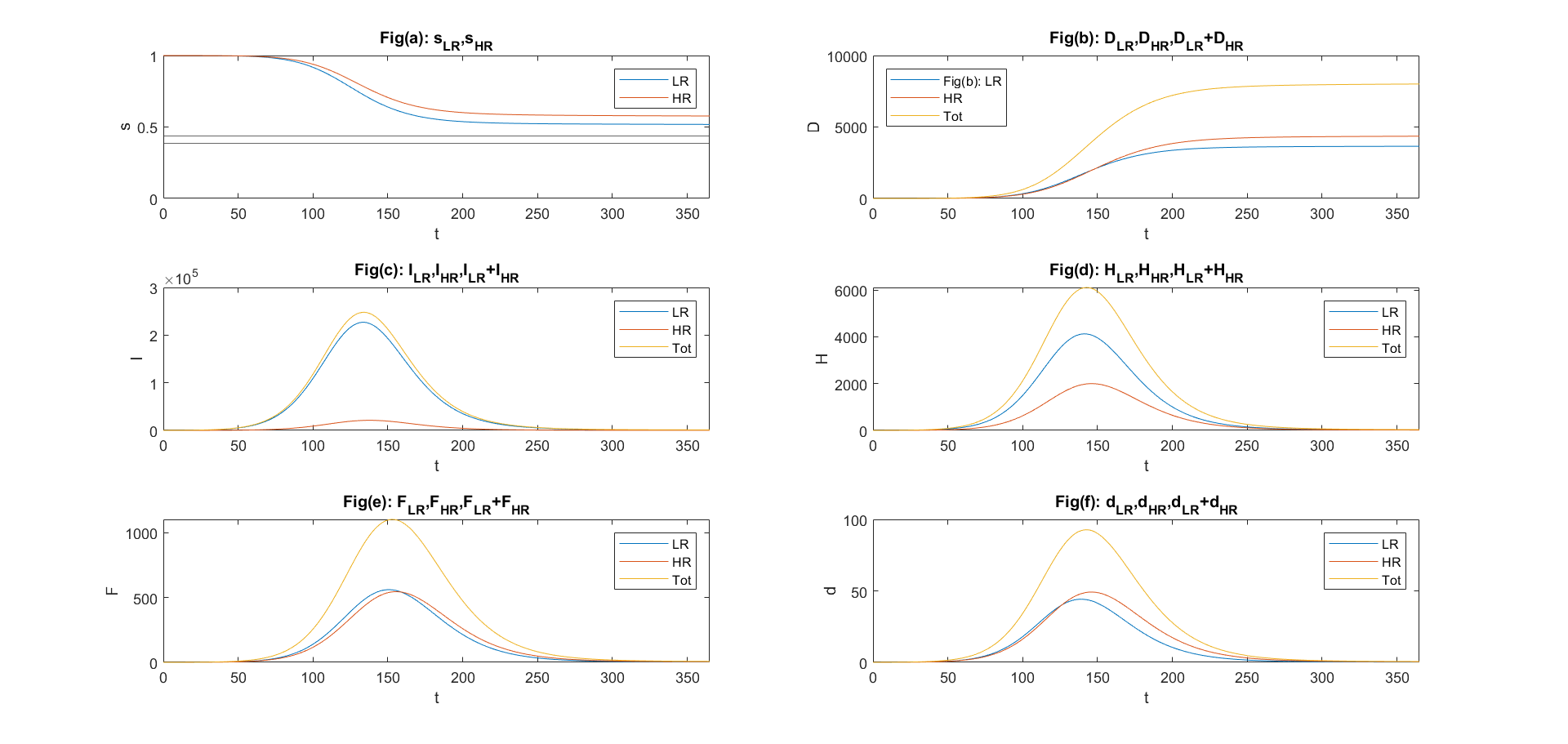}
\caption{{}Disease dynamics, in Sweden, sufficient number of ICU beds, with seasonality.}
\label{Fig3B}
\end{sidewaysfigure}

In Figure \ref{Fig3B}, we show the dynamic of the disease in Sweden
corresponding to the same parameters, but with seasonality accounted for: 
\begin{equation}
\begin{array}{c}
\rho _{1}=1.55(1+0.2\cos \left( \frac{2\pi t}{365}\right) ,\ \ \ \rho
_{2}=1.45(1+0.2\cos \left( \frac{2\pi t}{365}\right) .%
\end{array}
\label{Eq19b}
\end{equation}%
No quarantine is imposed. The number of ICU units is sufficient to satisfy
the entire demand.%
\begin{equation*}
\text{Figure \ref{Fig3B} near here.}
\end{equation*}

\begin{sidewaysfigure}[th]
{\center}%
\includegraphics[width=1.0\textwidth, height=1.0\textheight, angle=0]
{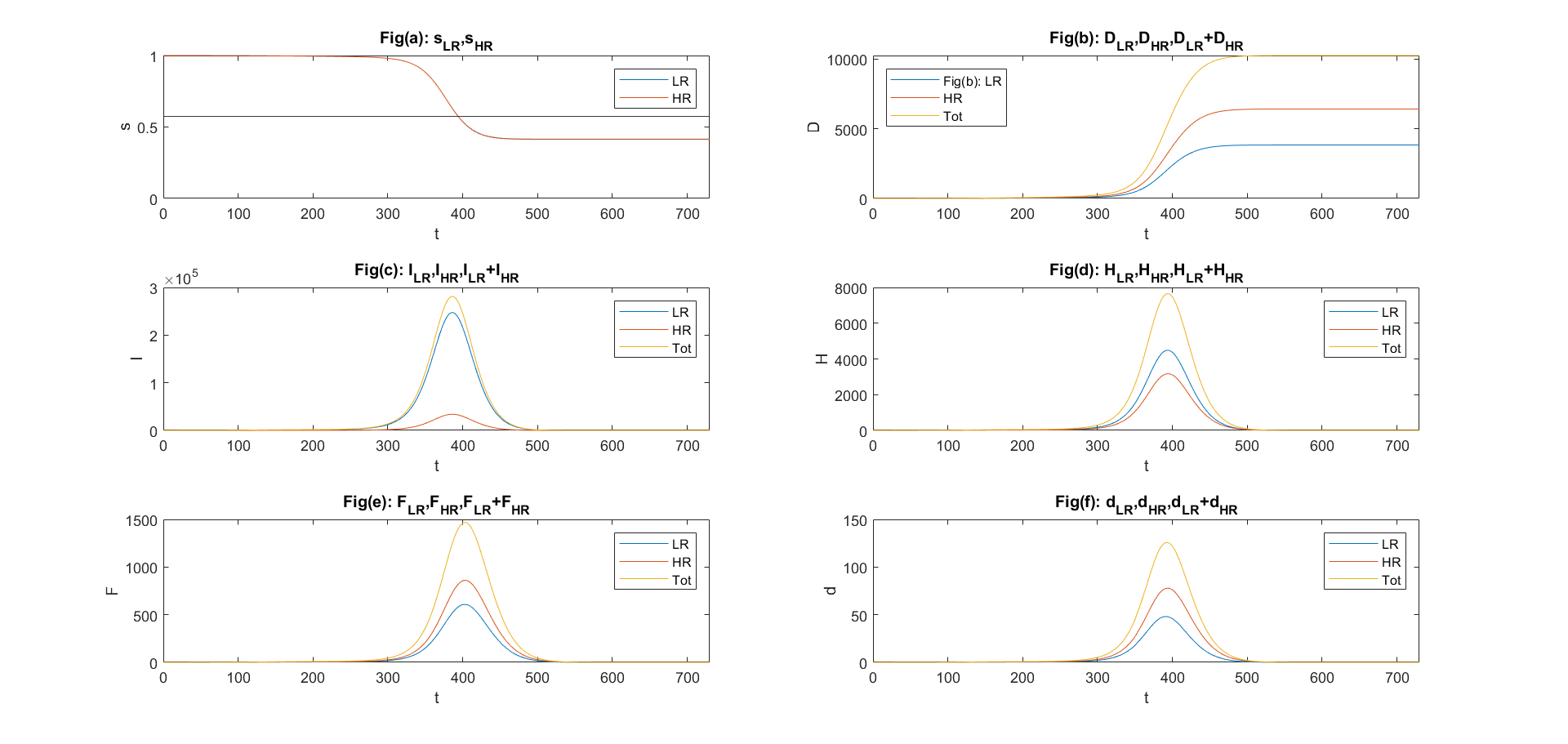}
\caption{{}Disease dynamics, in Israel, sufficient number of ICU beds.}
\label{Fig4}
\end{sidewaysfigure}

In Figure \ref{Fig4} we show the dynamic of the disease in Israel, with the
quarantine imposed after 30 days and kept for 45 days. The \textquotedblleft
natural\textquotedblright\ reproductive numbers for both groups are given by%
\begin{equation}
\begin{array}{c}
\rho _{k}\left( t\right) =1.3\left( 1+0.2\cos \left( \frac{2\pi t}{365}%
\right) \right) .%
\end{array}
\label{Eq20}
\end{equation}%
It is reduced by 30\% during the quarantine period. The number of ICU beds
is 1250 is sufficient to satisfy the demand.%
\begin{equation*}
\text{Figure \ref{Fig4} near here.}
\end{equation*}

\begin{sidewaysfigure}[th]
{\center}%
\includegraphics[width=1.0\textwidth, height=1.0\textheight, angle=0]
{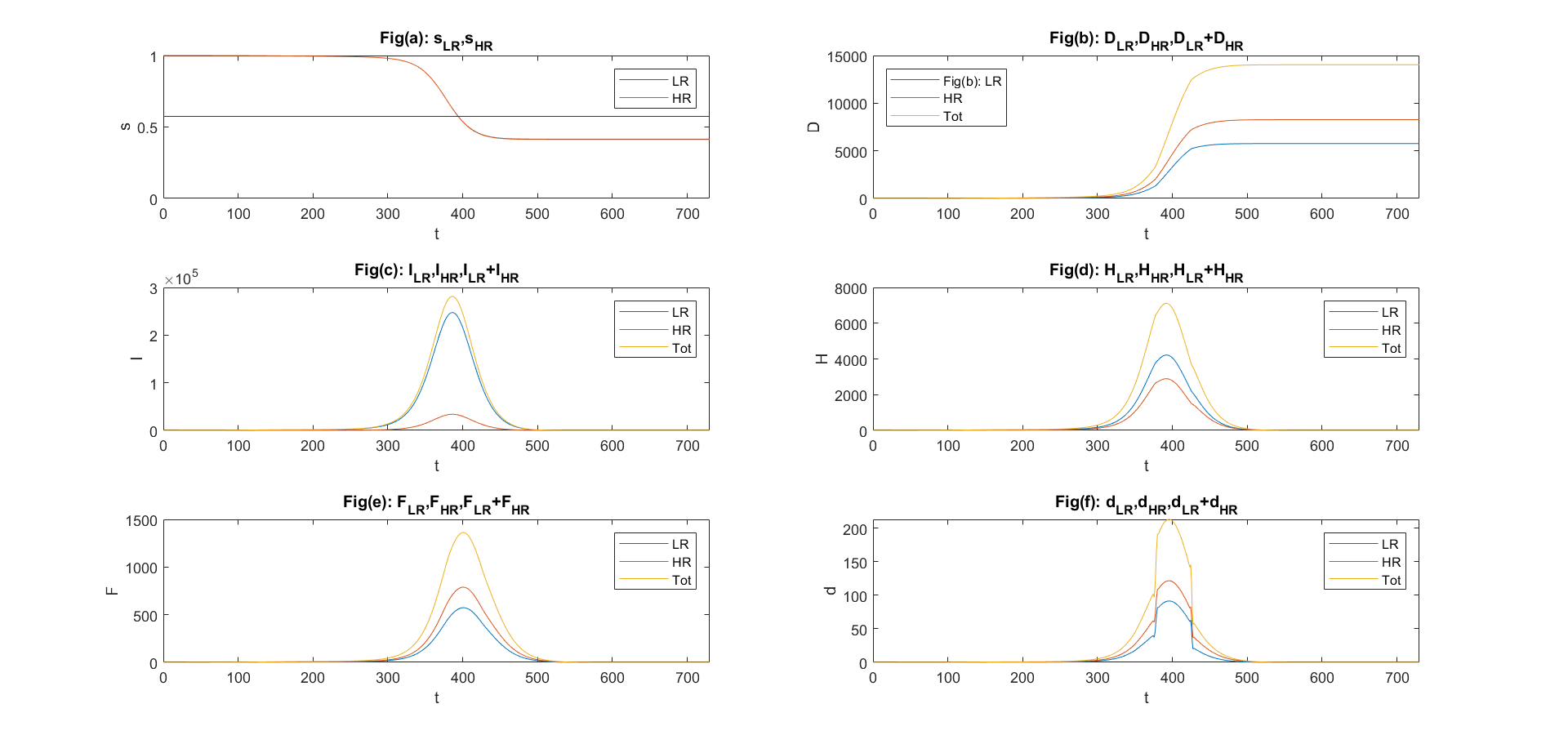}
\caption{{}Disease dynamics, in Israel, insufficient number of ICU beds.}
\label{Fig5}
\end{sidewaysfigure}

In Figure \ref{Fig5} we show the dynamic of the disease with the same
parameters as before, but with 600 ICU beds. The spike in mortality caused
by the ICU\ rationing is visible.%
\begin{equation*}
\text{Figure \ref{Fig5} near here.}
\end{equation*}

\begin{sidewaysfigure}[th]
{\center}%
\includegraphics[width=1.0\textwidth, height=1.0\textheight, angle=0]
{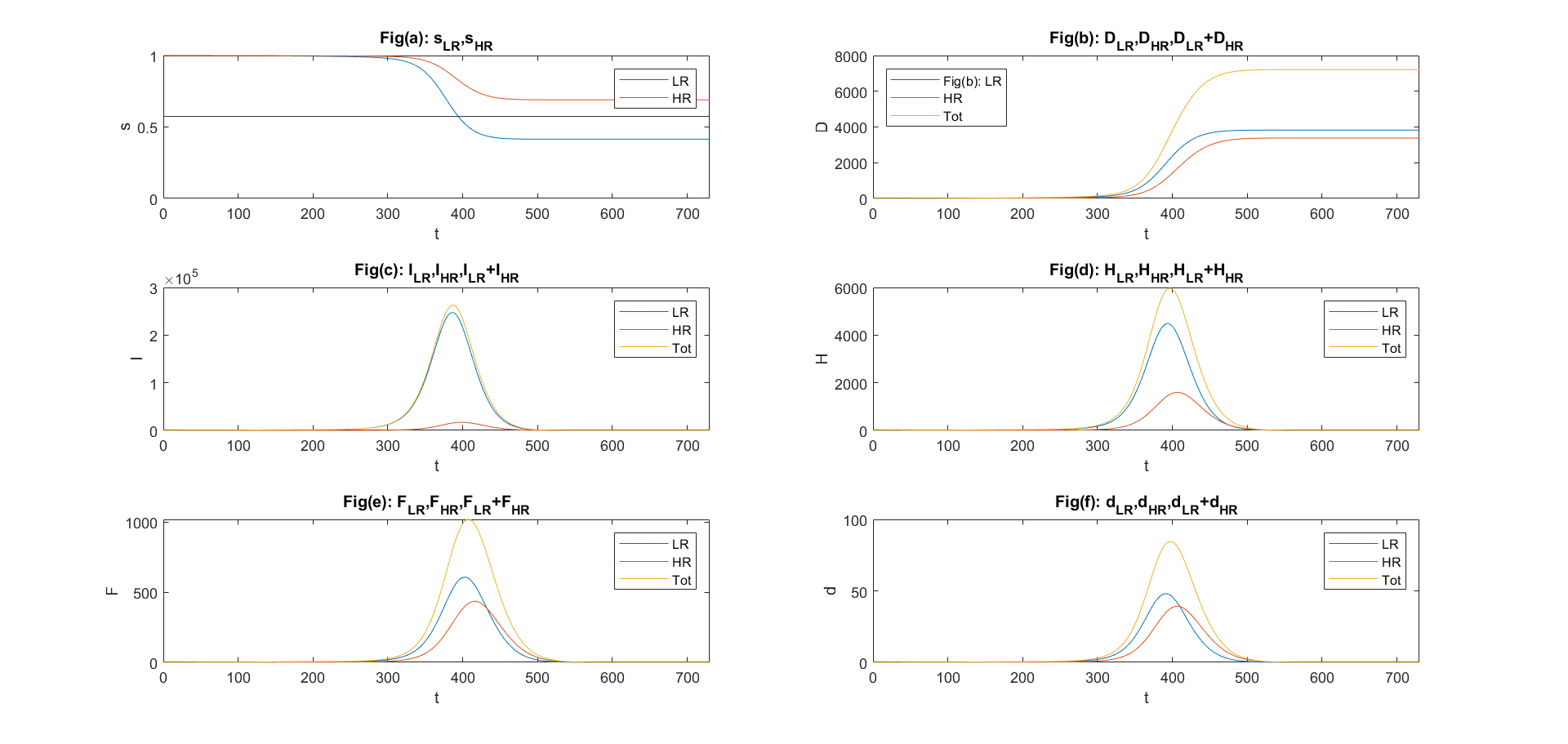}
\caption{{}Disease dynamics, in Israel, sufficient number of ICU beds. LR group released, HR group left in quarantine.}
\label{Fig6}
\end{sidewaysfigure}

\begin{sidewaysfigure}[th]
{\center}%
\includegraphics[width=1.0\textwidth, height=1.0\textheight, angle=0]
{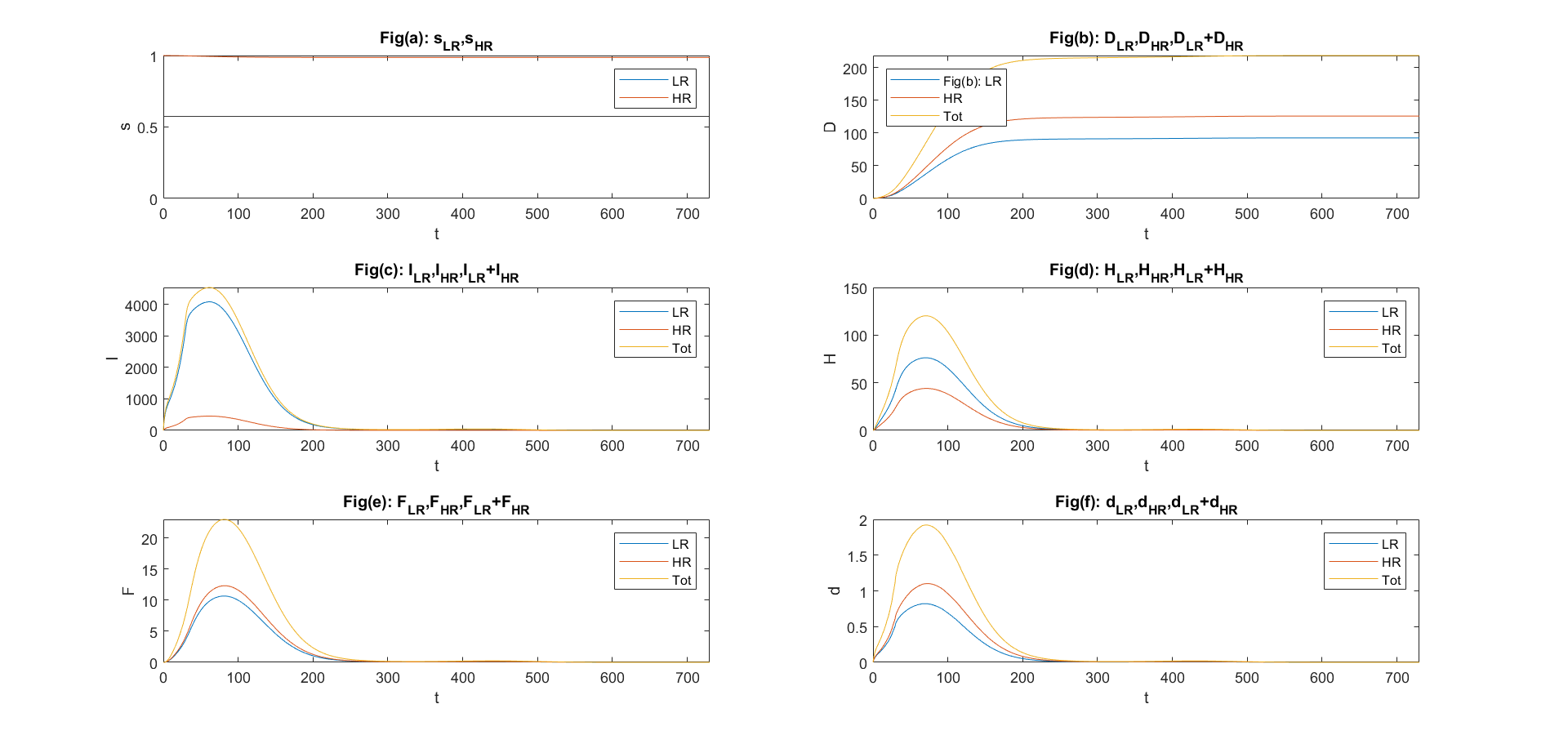}
\caption{{}Disease dynamics, in Israel, sufficient number of ICU beds. Both groups left in quarantine.}
\label{Fig7}
\end{sidewaysfigure}

\begin{sidewaysfigure}[th]
{\center}%
\includegraphics[width=1.0\textwidth, height=1.0\textheight, angle=0]
{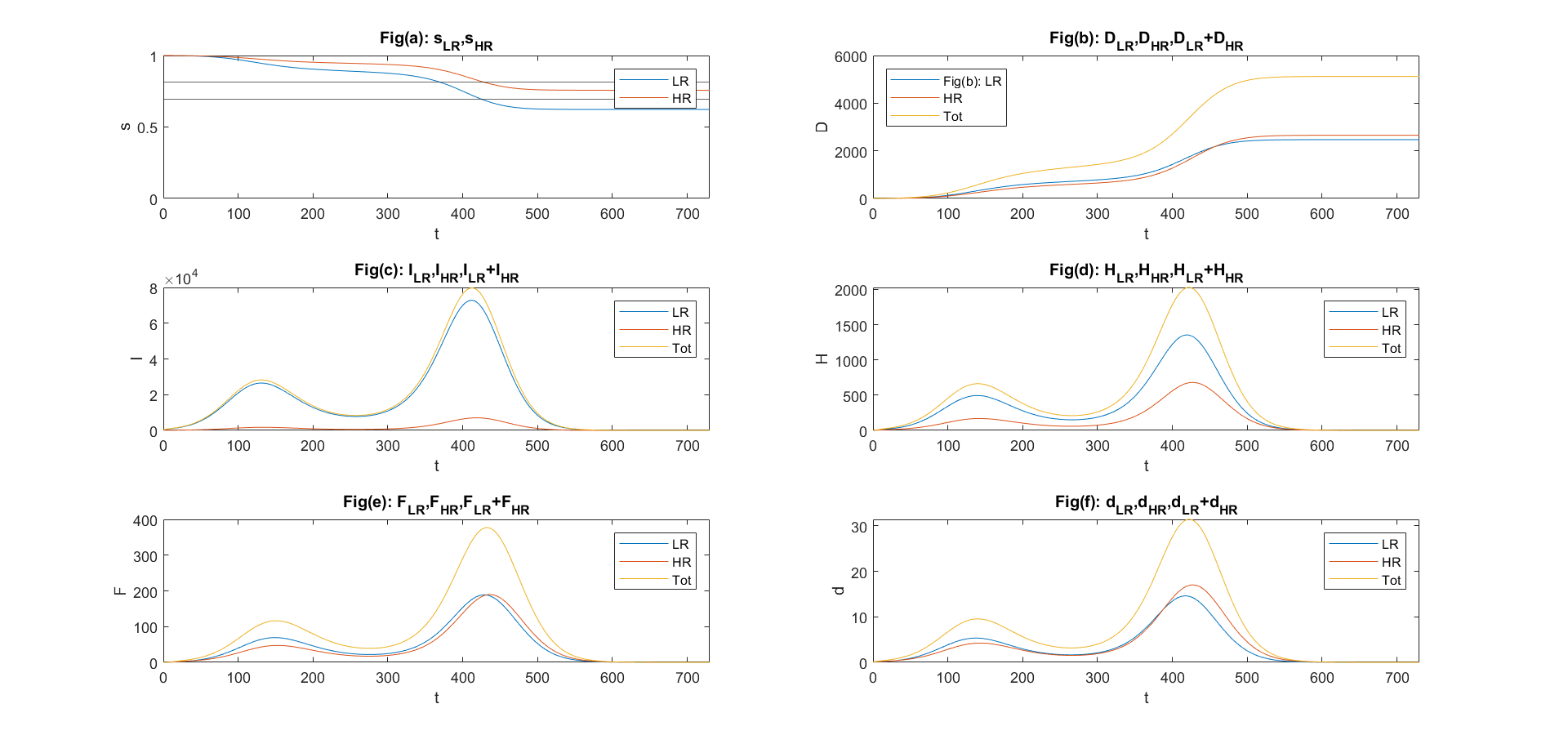}
\caption{{}Disease dynamics, in Israel, sufficient number of ICU beds. No quarantine at all, but strict mitigating measures are enforced. 
Seasonality is included.}
\label{Fig8}
\end{sidewaysfigure}

In Figures \ref{Fig6}, \ref{Fig7}, \ref{Fig8}, we analyze three approaches
to exiting the quarantine. In Figure \ref{Fig6}, we assume that the $HR$
group is left in the quarantine indefinitely, while the $LR$ group is not
restricted. Figure \ref{Fig7}, we assume that both groups are left in the
quarantine indefinitely (however improbable such an assumption is). Finally,
in the most promising \ref{Fig8}, we show what happens when both groups are
released from the quarantine. Still, strict disease mitigation is pursued,
allowing to reduce the reproductive numbers by 20\%.%
\begin{equation*}
\text{Figure \ref{Fig6} near here.}
\end{equation*}%
\begin{equation*}
\text{Figure \ref{Fig7} near here.}
\end{equation*}%
\begin{equation*}
\text{Figure \ref{Fig8} near here.}
\end{equation*}

\begin{sidewaysfigure}[th]
{\center}%
\includegraphics[width=1.0\textwidth, height=1.0\textheight, angle=0]
{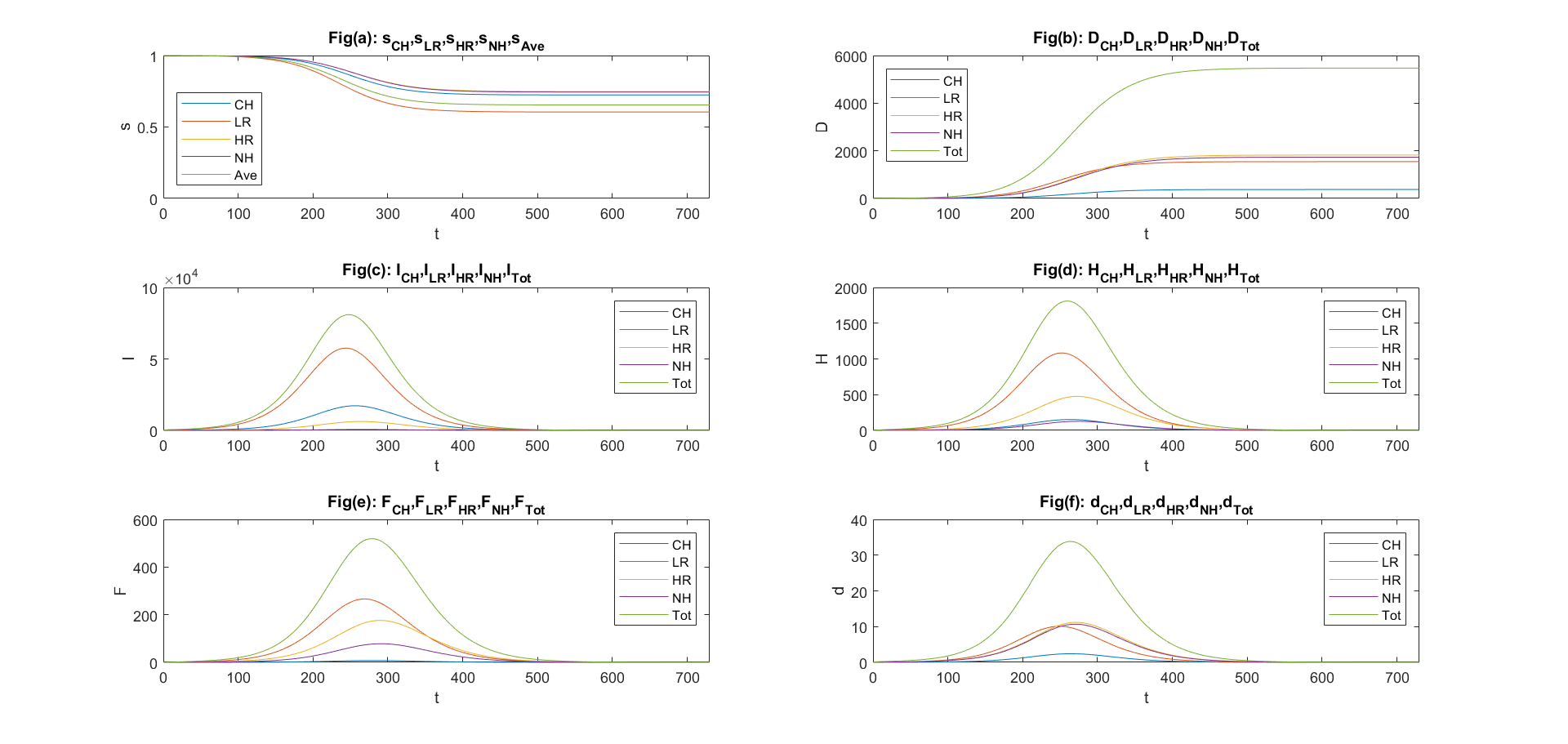}
\caption{{}Disease dynamics, in Israel, sufficient number of ICU beds. Four groups. $R_{0}=(1.1,1.3,1.1,1.1)$}
\label{Fig11}
\end{sidewaysfigure}

In Figure \ref{Fig11}, we show disease dynamics in Israel for the four-group
model, with sufficient number of ICU beds. We assume that $R_{0}$ is
constant, $R_{0}=\left( 1.1,1.3,1.1,1.1\right) $.%
\begin{equation*}
\text{Figure \ref{Fig11} near here.}
\end{equation*}

\begin{sidewaysfigure}[th]
{\center}%
\includegraphics[width=1.0\textwidth, height=1.0\textheight, angle=0]
{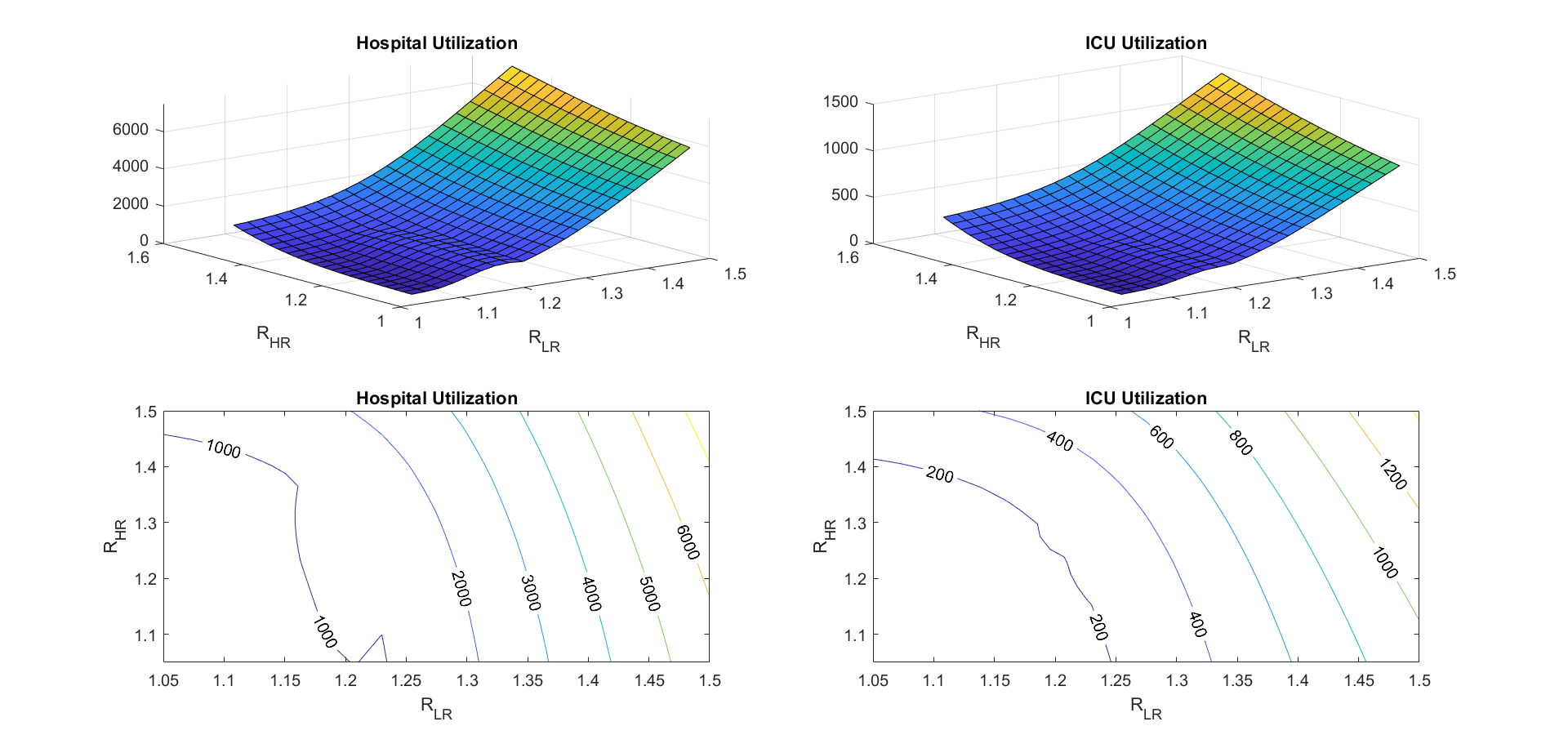}
\caption{{}Hospital and ICU consumption in Israel as functions of reproductive numbers. No quarantine. Other parameters are the same as in Figure \ref{Fig4}.}
\label{Fig9}
\end{sidewaysfigure}

\begin{sidewaysfigure}[th]
{\center}%
\includegraphics[width=1.0\textwidth, height=1.0\textheight, angle=0]
{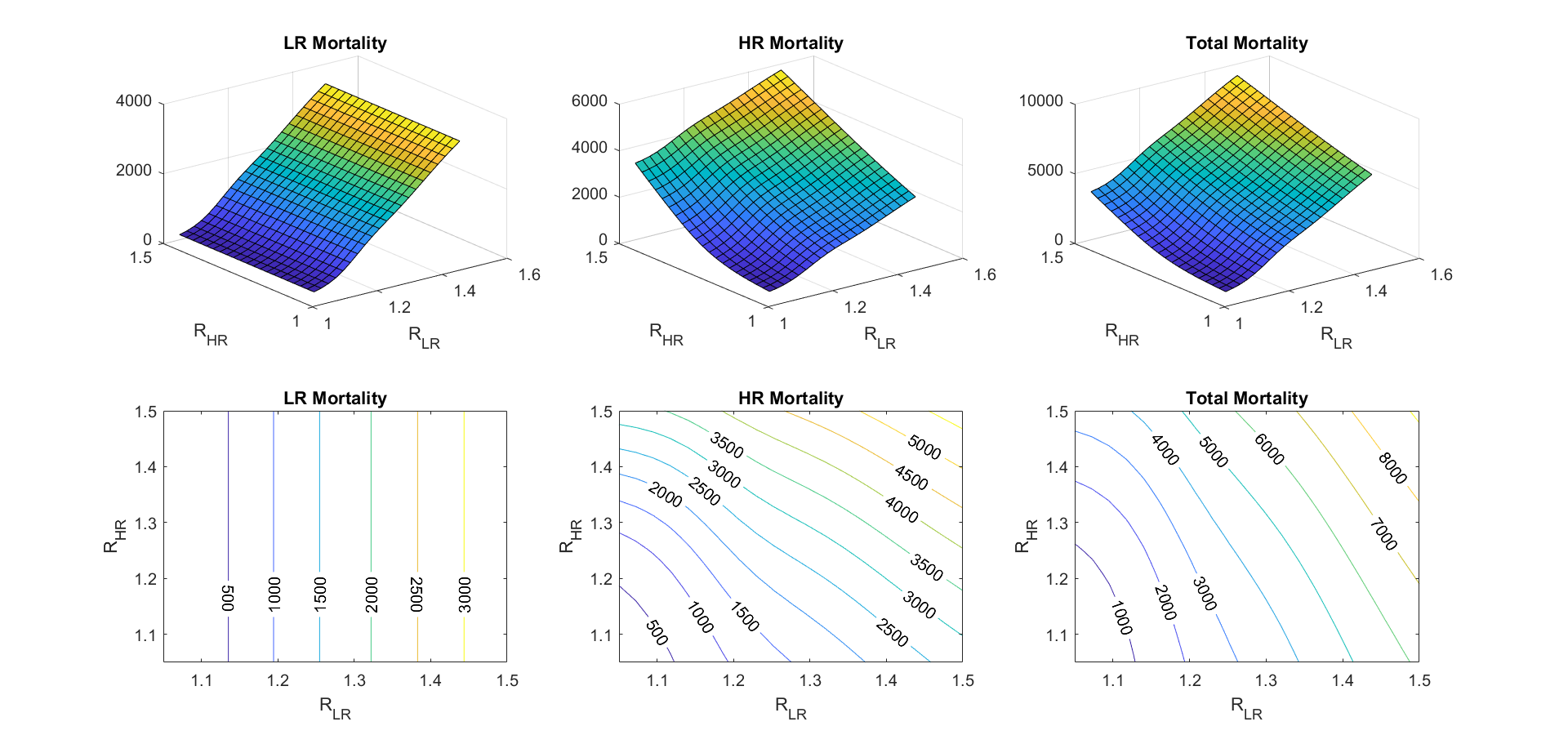}
\caption{{}Mortality in Israel as functions of reproductive numbers. No quarantine. Other parameters are the same as in Figure \ref{Fig4}.}
\label{Fig10}
\end{sidewaysfigure}

In Figures \ref{Fig9}, \ref{Fig10}, we show the hospital and ICU utilization
with the same parameters as in Figure \ref{Fig4}, for different choices of
the reproductive numbers. Similarly, in Figure \ref{Fig10}, we present the
total number of deaths.%
\begin{equation*}
\text{Figure \ref{Fig9} near here.}
\end{equation*}%
\begin{equation*}
\text{Figure \ref{Fig10} near here.}
\end{equation*}

\section{Conclusions\label{Conclusions}}

In this article, we develop a flexible multi-group framework for analyzing
the consequences of pandemics with various quarantine regimes. We compute
the extra capacity of the healthcare system, which is required to absorb all
ill and severely ill patients. Our analysis can easily include several
interacting groups. However, to achieve our goal in the most efficient way,
we focus on the case of two groups only - the high-risk and low-risk
populations - and examine the dynamics of the infection spread. We also
study the impact of transmission on the overall healthcare system and
demonstrate that the result varies quite dramatically depending on the
chosen strategies.

For a given level of parameters of the pandemic and the size of the health
care system, we define the dynamics of the disease, including the number of
infections, hospitalizations, ICU admissions, recoveries and deaths.

We extend the SEIR model to incorporate the two population groups. We
demonstrate that treating two groups as the same brings no additional
benefit (at least in the case of Israel) while causing significant economic
suffering.

Assuming that the population abides by pandemic public health
countermeasures, thereby effectively reducing transmission of the disease,
and given the health system capacity at or above the threshold, the country
can impose no quarantine at all. Examples include Singapore, Hong Kong,
Taiwan, and Sweden. None of them enforced any lockdown, which is consistent
with our results.

Social distancing reducing the reproductive number in big cities can be
enforced by making changes in the work hours and practicing staggering.
These changes naturally lessen the burden on public transportation during
rush hours and reduce crowding during lunch breaks. For example, even
offices and schools can operate in shifts, while weekends can become
floating.

Our analysis is useful in countries with high levels of compliance with
hygienic and social distance measures. If the capacity of the health system
is large enough, and the number of hospital or ICU beds is at or above the
threshold level, then no quarantine is required. If the number of ICU beds
is below the threshold, then quarantine of the $HR$ population only can
allow the country to slow the spread of disease and flatten the curve. In
the meantime, the $LR$ group can support the economy while maintaining
pandemic interventions.

\begin{acknowledgement}
We are deeply grateful to Prof. Matheus Grasselli, Dr. Marsha Lipton and
Prof. Marcos Lopez de Prado for extremely viable interactions.
\end{acknowledgement}

\begin{table}[tbp]
\begin{center}
\begin{tabular}{|c|c|c|c|c|c|c|c|c|c|}
\hline
& $\beta ^{\left( H\right) }$ & $\beta ^{\left( R\right) }$ & $\beta
^{\left( D\right) }$ & $\gamma ^{\left( H\right) }$ & $\gamma ^{\left(
R\right) }$ & $\gamma ^{\left( D\right) }$ & $\delta ^{\left( H\right) }$ & $%
\delta ^{\left( D\right) }$ & $\pi $ \\ \hline
$k=1$ & $1.0$ & $98.95$ & $0.05$ & $10.0$ & $89.0$ & $1.0$ & $80.0$ & $20.0$
& $90.0$ \\ \hline
$k=2$ & $5.0$ & $94.5$ & $0.5$ & $20.0$ & $78.0$ & $2.0$ & $60.0$ & $40.0$ & 
$80.0$ \\ \hline
\end{tabular}%
\end{center}
\caption{2-SEIR parameters in percent, $1\rightarrow LR$, $2\rightarrow HR$.}
\label{Tab1}
\end{table}

\begin{table}[tbp]
\begin{center}
\begin{tabular}{|c|c|c|c|c|c|c|c|c|c|}
\hline
& $\beta ^{\left( H\right) }$ & $\beta ^{\left( R\right) }$ & $\beta
^{\left( D\right) }$ & $\gamma ^{\left( H\right) }$ & $\gamma ^{\left(
R\right) }$ & $\gamma ^{\left( D\right) }$ & $\delta ^{\left( H\right) }$ & $%
\delta ^{\left( D\right) }$ & $\pi $ \\ \hline
$k=1$ & $0.5$ & $99.50$ & $0.0$ & $2.0$ & $87.0$ & $1.0$ & $95.0$ & $5.0$ & $%
50.0$ \\ \hline
$k=2$ & $1.0$ & $98.95$ & $0.05$ & $10.0$ & $89.0$ & $1.0$ & $90.0$ & $10.0$
& $90.0$ \\ \hline
$k=3$ & $4.0$ & $95.5$ & $0.5$ & $15.0$ & $84.0$ & $1.0$ & $70.0$ & $30.0$ & 
$80.0$ \\ \hline
$k=4$ & $10.0$ & $85.0$ & $5.0$ & $25.0$ & $72.0$ & $3.0$ & $50.0$ & $50.0$
& $80.0$ \\ \hline
\end{tabular}%
\end{center}
\caption{4-SEIR parameters in percent, $1\rightarrow C$, $2\rightarrow LR$, $%
3\rightarrow HR$, $4\rightarrow NH$.}
\label{Tab2}
\end{table}

\end{document}